\documentclass[11pt,reqno]{JHEP3}
\usepackage{amsmath}
\usepackage{amsfonts}
\usepackage{amssymb}

\title{Charges of twisted branes: the exceptional cases}  
\author{Stefan Fredenhagen, Matthias R.\ Gaberdiel and Thomas Mettler\\ 
Institut f{\"u}r Theoretische Physik, ETH-H{\"o}nggerberg\\
CH--8093 Z{\"u}rich, Switzerland\\ E-mail:
\email{stefan@itp.phys.ethz.ch}, \email{gaberdiel@itp.phys.ethz.ch},
\email{thomas.mettler@student.ethz.ch}} 
\abstract{The charges of the twisted D-branes for the two exceptional
cases (SO(8) with the triality automorphism and $E_6$ with charge
conjugation) are determined. To this end the corresponding NIM-reps are
expressed in terms of the fusion rules of the invariant
subalgebras. As expected the charge groups are found to agree with 
those characterising the untwisted branes.} 
\preprint{hep-th/0504007}
\begin{document}

\section{Introduction}
A lot of information about the dynamics of D-branes is encoded in
their charges. In particular, the D-brane charges constrain possible
decay processes, and thus play an important role in stability
considerations. There is evidence that these charges take values in
(twisted) K-theory 
\cite{Minasian:1997mm,Witten:1998cd,Moore:2003vf}. For D-branes on a  
simply connected group manifold $G$, the charge group is conjectured
to be the twisted K-theory $^{k+h^\vee}K (G)$ 
\cite{Kapustin:1999di,Bouwknegt:2000qt}, where the twist involves an
element of the third cohomology group $H^{3} (G,\mathbb{Z})$, the 
Wess-Zumino form of the underlying Wess-Zumino-Witten (WZW) model at
level $k$.

For all simple, simply connected Lie groups $G$, the twisted K-theory
has been computed in \cite{Braun:2003rd} (see also \cite{FHT,D})   
to be
\begin{equation}\label{Kgroup}
^{k+h^\vee}K (G)\ =\ \underbrace{\mathbb{Z}_{M(G,k)}\oplus \dotsb \oplus 
\mathbb{Z}_{M(G,k)}}_{2^{{\rm rk} (G)-1}}\  ,
\end{equation}
where $M(G,k)$ is the integer 
\begin{equation}
M(G,k)\  =\  \frac{k+h^\vee}{{\rm gcd}(k+h^\vee,L)} \ . 
\end{equation}
Here $h^\vee$ is the dual Coxeter number of the finite dimensional
Lie algebra $\bar{\mathfrak{g}}$, and $L$ only depends on $G$ (but not
on $k$). In fact, except for the case of $C_n$ that will not concern
us in this paper, $L$ is 
\begin{equation}
L \ =\  {\rm lcm} \{1,2,\ldots,h-1\} \ ,
\end{equation} 
where $h$ is the Coxeter number of $\bar{\mathfrak{g}}$. For
$\bar{\mathfrak{g}}=A_n$ this formula was derived in 
\cite{Fredenhagen:2000ei} (see also \cite{MMS}), while the formulae in
the other cases were checked numerically up to very high levels in 
\cite{Bouwknegt:2002bq}. For the classical Lie algebras and $G_2$ an
alternative expression for $M$ was also derived in \cite{D}.
\smallskip

These results should be compared with the charges that can be
determined directly in terms of the underlying conformal field
theory. The idea behind this approach is that brane configurations 
that are connected by RG flows should carry the same charge. These
constraints were used in \cite{AS} to determine the charge group of
su$(2)$. The constraint equations were generalised in
\cite{Fredenhagen:2000ei} to the branes 
$a\in \mathcal{B}_{k}^{\omega}$ of an arbitrary WZW model that 
preserve the full affine  symmetry algebra $\mathfrak{g}$ up to some
automorphism $\omega$. There it was argued that the charges
$q_a$ satisfy
\begin{equation}\label{chargerelation}
\dim (\lambda)\  q_{a}\ =\ \sum_{b\in \mathcal{B}_{k}^{\omega }} 
\mathcal{N}_{\lambda a}{}^{b}\ q_{b}  \ ,
\end{equation}
where $\lambda\in \mathcal{P}_{k}^{+} (\bar{\mathfrak{g}})$ is a
dominant highest-weight representation of the affine Lie algebra
$\mathfrak{g}$ at level $k$, $\dim(\lambda)$ is the Weyl-dimension of
the corresponding representation of the horizontal subalgebra
$\bar{\mathfrak{g}}$, and $\mathcal{N}_{\lambda a}{}^{b}$ are the
NIM-rep coefficients appearing in the Cardy analysis. In this paper we
shall ignore the low level ($k=1,2$) subtleties discussed in
\cite{Bouwknegt:2002bq} and assume that $k$ is sufficiently big
($k\geq 3$).

For the trivial automorphism ($\omega =\text{id}$), the branes can be
labelled by dominant highest weights of $\mathfrak{g}$,
$\mathcal{B}_{k}^{\text{id}}\cong \mathcal{P}_{k}^{+}
(\bar{\mathfrak{g}})$. In this case, the
constraints~\eqref{chargerelation} were evaluated in
\cite{Fredenhagen:2000ei,Bouwknegt:2002bq}. The charges are given (up
to rescalings) by the Weyl-dimensions of the corresponding
representations, $q_{\lambda}=\dim (\lambda)$, and the charge is
conserved only modulo $M(G,k)$. Thus, the untwisted branes account
for one summand $\mathbb{Z}_{M(G,k)}$ of the K-group~\eqref{Kgroup}. 

For nontrivial outer automorphisms, a similar analysis was carried
through in \cite{Gaberdiel:2003kv}. Here, the D-branes are
parametrised by $\omega$-twisted highest weight representations $a$ of
$\mathfrak{g}_{k}$ \cite{Birke:1999ik,Fuchs:2000vg,Gaberdiel:2002qa},
and the NIM-rep coefficients are given by twisted fusion rules
\cite{Gaberdiel:2002qa}. The twisted representations can be
identified with representations of the 
invariant subalgebra $\bar{\mathfrak{g}}^{\omega}$ consisting of
$\omega$-invariant elements of $\bar{\mathfrak{g}}$, and we can view
$\mathcal{B}_{k}^{\omega}$ as a subset of
$\mathcal{P}_{k'}^{+} (\bar{\mathfrak{g}}^{\omega })$, where 
$k'=k+ h^\vee(\bar{\mathfrak{g}}) 
-h^\vee(\bar{\mathfrak{g}}^{\omega})$. 
It was found that the charge $q_a$ of $a\in \mathcal{B}_{k}^{\omega}$ 
is again (up to rescalings) given by the Weyl dimension\footnote{A
similar proposal was made in~\cite{Alekseev:2002rj} 
based on an analysis for large level.} of the
representation of $\bar{\mathfrak{g}}^{\omega}$, $q_{a}=\dim (a)$, and
that the charge identities are only satisfied modulo $M(G,k)$. Thus
each such class of twisted D-branes accounts for another summand 
$\mathbb{Z}_{M(G,k)}$ of the charge group.  Since the number of
automorphisms does not grow with the level, these 
constructions do not in general account for all the charges of
(\ref{Kgroup}); for the case of the $A_n$ series, a proposal for the
D-branes that may carry the remaining charges was made in 
\cite{Gaberdiel:2004hs,Gaberdiel:2004za} (see also
\cite{Maldacena:2001xj}). 

The analysis of \cite{Gaberdiel:2003kv} was only done for all order-2 
automorphisms of the classical Lie groups.  There exist two
`exceptional' automorphisms, namely the order-3 automorphism of $D_4$
(triality), and the order-2 automorphism of $E_6$ (charge
conjugation). These two cases are the subject of this paper. We will
find again that with the charge assignment $q_{a}=\dim (a)$, the charge
identities are satisfied modulo $M(G,k)$. Thus each such class of
D-branes accounts for another summand of the charge
group.\footnote{For the case of $D_4$, there are in total five
`twisted' classes of branes that are associated to $\omega$,
$\omega^2$, $C$, $\omega C$ and $\omega^2 C$, where $C$ denotes charge
conjugation. The corresponding NIM-reps are all closely related to the
one discussed in this paper, or the charge conjugation NIM-rep
discussed in \cite{Gaberdiel:2003kv} (see
\cite{Gaberdiel:2002qa}). The arguments given here, together with the 
results of \cite{Gaberdiel:2003kv} therefore imply that these five
twisted classes of D-branes account for five summands in (\ref{Kgroup}).}
\smallskip

The plan of the paper is as follows. In the remainder of this section
we shall explain the main steps in proving these results that are
common to both cases. The details of the analysis for the case of the 
triality automorphism of $D_{4}$ (whose invariant subalgebra is
$G_{2}$) is given in section~2. The corresponding analysis for the
charge conjugation automorphism of $E_6$ (whose invariant subalgebra
is $F_4$) is given in section~3.

\subsection{Some notation and a sketch of the proof}

We begin by briefly introducing some notation. The $\omega$-twisted
D-branes are characterised by the gluing conditions 
\begin{equation}\label{gluing}
\left( J^a_n + \omega(\bar{J}^a_{-n}) \right)\, 
|\!| a\rangle\!\rangle \ =\  0 \  ,
\end{equation}
where $J^a_n$ are the generators of $\mathfrak{g}$. Every boundary
state can be written in terms of the $\omega$-twisted Ishibashi states
\begin{equation}\label{expa}
|\!| a\rangle\!\rangle \ =\  
\sum_{\mu \in {\cal E}_k^\omega} \psi_{a\mu}\, 
|\mu\rangle\!\rangle^\omega \  ,
\end{equation}
where $|\mu\rangle\!\rangle^\omega$ is the (up to normalisation)
unique state satisfying (\ref{gluing}) 
in the sector ${\cal H}_\mu \otimes \bar{\cal H}_{\mu^\ast}$. The sum
in (\ref{expa}) runs over the so-called exponents that consist of the
weights $\mu\in {\cal P}^{+}_k(\bar{\mathfrak{g}})$ that are invariant
under $\omega$. The NIM-rep coefficients are determined by the
Verlinde-like formula  
\begin{equation}\label{NIMrepgen}
{\cal N}_{\lambda a}{}^{b} \ =\  \sum_{\mu \in {\cal E}_k^\omega} 
\frac{ \psi_{b\mu}^\ast\, S_{\lambda \mu}\, \psi_{a\mu}}{S_{0\mu}} \ ;
\end{equation}
they define a non-negative integer matrix representation (NIM-rep) of
the fusion rule algebra. (For a brief review of these matters see for
example \cite{Gaberdiel:2003kv} and \cite{BPPZ}.) 

It is clear on general grounds (see \cite{Gaberdiel:2003kv}) that for
any charge assignment $q_a$ for $a\in \mathcal{B}_{k}^{\omega}$, the
charge identity (\ref{chargerelation}) can at most be satisfied modulo
$M(G,k)$. Our strategy will therefore be to construct a solution that
solves (\ref{chargerelation}) modulo $M(G,k)$. This solution is again
given by $q_a = \dim(a)$. Furthermore,
we can show that this solution of the charge equation is unique (up to
trivial rescalings).

Our arguments will depend on the particularities of the two cases, but
the general strategy is the same. The key observation of our analysis 
in both cases is a relation of the form
\begin{equation}\label{NIMrep}
\mathcal{N}_{\lambda a}{}^{b}\ =\ \sum_{\gamma,i}
\varphi_{\lambda}{}^{\gamma}\,\varepsilon_{i}\, 
N_{\gamma a}{}^{\rho_{i}(b)} 
\ \ 
\end{equation}
that expresses the NIM-rep coefficients $\mathcal{N}_{\lambda a}{}^{b}$
in terms of the fusion rules $N_{\gamma a}{}^{\rho_{i} (b)}$ of  the
affine algebra corresponding to $\bar{\mathfrak{g}}^\omega$. Here
$\varphi_{\lambda}{}^{\gamma}$ is the 
branching coefficient which denotes how often the representation
$\gamma$ of $\bar{\mathfrak{g}}^{\omega}$ appears in the restriction
of the representation $\lambda$ to $\bar{\mathfrak{g}}$. The
$\rho_{i}$ are maps 
$\rho_{i}:\mathcal{B}_k^{\omega}\to 
\mathcal{P}_{k'}^{+}(\bar{\mathfrak{g}}^{\omega})$ 
and $\varepsilon_{i}$ is a sign attributed to the map
$\rho_{i}$. Furthermore $k'$ is defined as before, 
$k'=k+h^\vee(\bar{\mathfrak{g}}) - h^\vee(\bar{\mathfrak{g}}^\omega)$. 
In the cases studied in \cite{Gaberdiel:2003kv} analogous formulae
for the NIM-rep coefficients were used for which the $\rho_i$ could be
expressed in terms of simple currents. In the current context where
the invariant algebras are $G_2$ and $F_4$, such simple currents do
not exist. Nevertheless it is possible to find such maps  
$\rho_{i}$ (see (\ref{eq:aff1}) and (\ref{eq:aff2}) below for the
specific formulae) that 'mimic' the action of the simple currents.   

\noindent The different maps $\rho_i$ have disjoint images, and we can
write  
\begin{equation}
{\cal P}_{k'}^{+}(\bar{\mathfrak{g}}^{\omega}) \ =\  \bigcup_i 
\rho_i\left(\mathcal{B}_k^{\omega}\right)  \ \ \cup \ {\mathcal R}_k 
\ ,
\end{equation}
where ${\mathcal R}_k$ denotes the remainder. The second key
ingredient in our proof are the relations 
\begin{alignat}{2}
\label{keydim1}
\dim (\rho_{i} (a)) \ & = \   \varepsilon_{i}\dim (a) &\qquad\quad   &
a\in \mathcal{B}_k^{\omega}\\ 
\label{keydim2}
\dim (b)\  & = \  0 &&  b\in {\mathcal R}_k \  .
\end{alignat}
Both of these identities hold modulo $M(G^\omega,k')$. Finally we
observe by explicit inspection of the above formulae for $M(G,k)$
that in the two cases of interest 
\begin{equation}
M(G,k) \ =\  M(G^\omega,k') \  .
\end{equation} 
This then allows us to reduce the proof of the charge identities for
the twisted D-branes of $G$ to that of the untwisted D-branes of
$G^\omega$. In fact, the argument is simply 
\begin{align}
\sum_{b \in \mathcal{B}_{k}^\omega}
\mathcal{N}_{\lambda a}^{\phantom{\lambda a}b}\,
\textrm{dim}_{G^\omega}(b)\ & =\ \sum_{b \in
  \mathcal{B}_{k}^{\omega}}\sum_{i}\sum_{\gamma}
\varepsilon_{i}\,\varphi^{\phantom{\lambda}\gamma}_{\lambda}
N_{\gamma a}^{\phantom{\lambda a}\rho_{i}(b)}\,
\textrm{dim}_{G^\omega}(b) &  \nonumber\\ 
& =\ \sum_{b \in \mathcal{B}_{k}^{\omega}}
\sum_{i}\sum_{\gamma}\varphi^{\phantom{\lambda}\gamma}_{\lambda}
N_{\gamma a}^{\phantom{\lambda a}\rho_{i}(b)}\,
\textrm{dim}_{G^\omega}(\rho_{i}(b)) 
& {\rm mod}\  M(G,k) \nonumber\\
& =\ \sum_{\gamma}\varphi^{\phantom{\lambda}\gamma}_{\lambda}
\sum_{b \in \mathcal{P}_{k'}^{+}(\bar{\mathfrak{g}}^\omega)} 
N_{\gamma a}^{\phantom{\lambda a}b}\, 
\textrm{dim}_{G^\omega}(b)
& {\rm mod}\  M(G,k) \nonumber\\ 
&=\ \sum_{\gamma}\varphi^{\phantom{\lambda}\gamma}_{\lambda}\,
\textrm{dim}_{G^\omega}(\gamma)\,\textrm{dim}_{G^\omega}(a) 
& {\rm mod}\  M(G,k) \nonumber \\
& =\ \textrm{dim}_{G}(\lambda)\,\textrm{dim}_{G^\omega}(a) \  . &
\end{align}
In the following two sections we shall give the details for how to
define the maps $\rho_i$, and prove the various statements above. We
shall also be able to show that our charge solution is unique up to
trivial rescalings.

\section{The $\boldsymbol{D_{4}}$ case with triality} 

In the $D_{4}$ case the relevant automorphism $\omega$ is triality
which maps the Dynkin labels
$\mu=(\mu_{0};\mu_{1},\mu_{2},\mu_{3},\mu_{4})$ to
$(\mu_{0};\mu_{4},\mu_{2},\mu_{1},\mu_{3})$. The set of exponents
labelling the $\omega $-twisted Ishibashi states is therefore 
\begin{equation}
\mathcal{E}^{\omega}_{k}=\{
(\mu_{0};\mu_{1},\mu_{2},\mu_{1},\mu_{1})\in
\mathbb{N}_{0}^{5}\,\vert\,\mu_{0}+3\mu_{1}+2\mu_{2}=k\}\ .
\end{equation}
The $\omega$-twisted boundary states are labelled by the level $k$ 
integrable highest weights of the twisted Lie algebra
$\mathfrak{g}^{\omega}=D_{4}^{(3)}$, which are 
$\mathcal{B}_{k}^\omega=\{(b_{0};b_{1},b_{2}) \in \mathbb{N}^{3}_{0} 
\,\vert\, b_{0}+2b_{1}+3b_{2}=k\}$. The states of lowest conformal
weight of these representations form irreducible representations of
the invariant subalgebra $\bar{\mathfrak{g}}^{\omega}=G_2$ with
highest weights $(b_{1},b_{2})$. For this reason we propose that the
corresponding D-brane charge is the Weyl dimension of these
irreducible representations, {\it i.e.}
\begin{equation} \label{eq:choice}
q_{b}\ =\ \textrm{dim}_{G_2}(b_{1},b_2)\ =\ \textrm{dim}_{G_2}(b)\ .
\end{equation}
In this section we shall prove that \eqref{eq:choice} solves 
the charge constraint
\begin{equation} \label{eq:main}
\textrm{dim}_{D_4}(\lambda)\,q_a\ =\ \sum_{b\in
\mathcal{B}_{k}^{\omega}}\mathcal{N}_{\lambda a}^{\phantom{\lambda a}b} \,
q_b 
\end{equation}
modulo $M(G,k)$ and that this solution is unique (up to rescalings).
\medskip

\subsection{The solution} 
To show that~\eqref{eq:choice} indeed solves the charge
constraint~\eqref{eq:main} we trace the problem back to the case of
untwisted branes in $G_{2}$. So we need to express 'twisted $D_{4}$
data' by 'untwisted $G_{2}$ data'. We first note, as already
mentioned in section~1.1, that the integer $M$ for $G_{2}$ at level 
$k+2$ equals the integer for $D_{4}$ at level $k$, 
\begin{equation}
M(D_{4},k)\ =\  M(G_{2},k+2)\  =\  \frac{k+6}{{\rm gcd}(k+6,60)}\ .
\end{equation}
As is also explained there, the key result (\ref{NIMrep}) that we need
to prove expresses the NIM-rep $\mathcal{N}$ of $D_4$ in terms of the
fusion rules of $G_{2}$. The first step in providing such a relation
is the identification of the $D_{4}$ $\psi$-matrix at level $k$   
with the (rescaled) $S$-matrix of $G_{2}$ at level
$k+2$ (in the following we shall denote the
$S$-matrix of $G_2$ by $S'$ in order to distinguish it from the
$S$-matrix of $D_4$)\footnote{This relation was already noted in
\cite{Gaberdiel:2002qa}.} 
\begin{equation}\label{eq:psiSident}
\psi_{b\mu}\ =\ \sqrt{3}\,S^{\prime}_{b\tilde{\mu}}\ ,
\end{equation}
where $\tilde{\mu}$ is defined by 
\begin{equation}\label{eq:idd}
\mu \ \mapsto\  \tilde{\mu} = (\mu_{0};3\mu_{1}+2,\mu_{2}) \ .
\end{equation}
Note that if $\mu \in {\cal E}^\omega_k$, then 
$\tilde{\mu}\in\mathcal{P}^{+}_{k+2}(G_{2})\equiv
\mathcal{P}_{k+2}=\{(\tilde\mu_{1},\tilde\mu_{2})\in
\mathbb{N}_{0}^{2}\,\vert\,\tilde\mu_{1}+2\tilde\mu_{2}\leq k+2\}$. 
The identity (\ref{eq:psiSident}) can be proven as follows. Define
$\kappa =k+6$ and  $c(x)=\cos\big(\frac{2\pi x}{3\kappa} \big)$. The
$\psi$-matrix is given by (see~\cite{Gaberdiel:2002qa}),
\begin{align}\label{eq:psi}
\nonumber
\psi_{b\mu}\  = \ \frac{2}{\kappa} \big( 
&c (pp'+2pq'+2qp'+qq') + c (2pp'+pq'+qp'-qq')\\
\nonumber
&+c (-pp'+pq'+qp'+2qq') - c (2pp'+pq'+qp'+2qq')\\
& - c (pp'+2pq'-qp'+qq') -c (pp'-pq'+2qp'+qq')
\big) \ , 
\end{align}
where $p=b_{1}+b_{2}+2$, $q=b_{2}+1$ and 
$p'=3\mu_{1}+\mu_{2}+4$, $q'=\mu_{2}+1$. 
On the other hand, if we define $m=\lambda_{1}+\lambda_{2}+2$,
$n=\lambda_{2}+1$, $m^{\prime}=\nu_{1}+\nu_{2}+2$ and
$n^{\prime}=\nu_{2}+1$, then the $S$-matrix of $G_{2}$ at level $k+2$
is~\cite{Gannon:2001py} 
\begin{align} \label{eq:g2}
\nonumber
S'_{\lambda\nu}\ =\ \frac{-2}{\sqrt{3}\kappa}& \big(
c(2mm^{\prime}+mn^{\prime}+nm^{\prime}+2nn^{\prime})
+c(-mm^{\prime}-2mn^{\prime}-nn^{\prime}+nm^{\prime})\\
& +c(-mm^{\prime}+mn^{\prime}-2nm^{\prime}-nn^{\prime})
-c(-mm^{\prime}-2mn^{\prime}-2nm^{\prime}-nn^{\prime})\nonumber\\
&
-c(2mm^{\prime}+mn^{\prime}+nm^{\prime}-nn^{\prime})
-c(-mm^{\prime}+mn^{\prime}+nm^{\prime}+2nn^{\prime})\big)\ .
\end{align}
For  \eqref{eq:g2} we also use the abbreviated notation 
\begin{equation}\label{abb}
S'_{\lambda\nu}\ =\ \frac{-2}{\sqrt{3}\kappa}\{c(u_{1})
+c(u_{2})+c(u_{3})-c(u_{4})-c(u_{5})-c(u_{6})\} \ .
\end{equation}
By comparing~\eqref{eq:psi} and~\eqref{eq:g2} one then easily
proves~\eqref{eq:psiSident}.

Next we observe from (\ref{NIMrepgen}) that in order to obtain 
fusion matrices of $G_{2}$ we also need to express the quotient 
$\frac{S_{\lambda\mu}}{S_{0 \mu}}$ in terms of $G_{2}$ $S$-matrices.
The relevant relation is 
\begin{equation}\label{Squot}
\frac{S_{\lambda\mu}}{S_{0\mu}}
\ =\ \sum_{\gamma}\varphi^{\phantom{\lambda}\gamma}_{\lambda}\ 
\frac{S^{\prime}_{\gamma\tilde{\mu}}}{S^{\prime}_{0\tilde{\mu}}}
\ .
\end{equation}
Here, $\varphi^{\phantom{\lambda}\gamma}_{\lambda}$ denotes the
$D_{4}\supset G_{2}$ branching rules; the most important ones are
\begin{equation}\label{D4branching}
(1,0,0,0)\ \to\  (1,0)\oplus (0,0)\quad \text{and}\quad 
(0,1,0,0)\ \to\  (0,1)\oplus
(1,0)\oplus (1,0) \ .
\end{equation}
The easiest way to prove~\eqref{Squot} is to consider the explicit
expressions for the fundamental representations of $D_{4}$.

Taking all of this together we can now write the $D_{4}$ NIM-rep as
\begin{equation} \label{eq:nim}
\mathcal{N}_{\lambda a}^{\phantom{\lambda a}b}\ 
=\ \sum_{\mu \in \mathcal{E}_{k}^{\omega}}
\psi_{a\mu}\,\psi_{b\mu}^{*}\,
\frac{S_{\lambda\mu}}{S_{0\mu}}
\ =\ 
3\sum_{\gamma}\varphi_{\lambda}{}^{\gamma} \,
\sum_{\mu \in \mathcal{E}_{k}^{\omega}}
S_{a\tilde{\mu}}^{\prime}\,
S_{b\tilde{\mu}}^{\prime*}\,
\frac{S'_{\gamma\tilde{\mu}}}{S'_{0\tilde{\mu}}} \ .
\end{equation}
Although this formula reminds one of the Verlinde formula, the last
sum still does not give the $G_{2}$-fusion rules as the range of
summation for $\tilde{\mu}$ is only a subset of 
$\mathcal{P}_{k+2}$.  

To resolve this problem we introduce the affine mappings 
\begin{equation} \label{eq:aff1}
\begin{array}{l}
\rho_{0}({b})\ =\ (b_{1},b_{2})\\
\rho_{1}({b})\ =\ (k-2b_{1}-3b_{2},1+b_{1}+b_{2})\\
\rho_{2}({b})\ =\ (k+1-b_{1}-3b_{2},b_{2})\\
\end{array}
\end{equation}
which map the set $\mathcal{B}_{k}^\omega$ of boundary states to
disjoint subsets of $\mathcal{P}_{k+2}=\mathcal{P}_{k+2}^+(G_2)$.
They have the crucial property
\begin{equation} \label{eq:killer1}
S^{\prime}_{\rho_{0} (b)\, \nu}
+S^{\prime}_{\rho_{1} (b)\, \nu}
-S^{\prime}_{\rho_{2}(b)\, \nu}
\ =\  \left\{ \begin{array}{ll} 3 \, S^{\prime}_{b\, \nu} \qquad & 
\textrm{if $\nu_{1}=2$ mod $3$}\\ 0 \qquad & \textrm{otherwise,} 
\end{array}
\right. 
\end{equation}
where $b \in \mathcal{B}_{k}^\omega$ and 
$\nu \in \mathcal{P}_{k+2}$. This follows from the fact that the 
left hand side can be written as  
\begin{align}
\frac{-\sqrt{3} \kappa}{2}\big(S^{\prime}_{b\, \nu}
+S^{\prime}_{\rho_1(b)\, \nu}
-S^{\prime}_{\rho_2(b)\, \nu}\big) \  
=\ &\big(1+\cos(v_{1})+\cos(v_{2})\big)\big(c(u_{1})-c(u_{4})\big)
\nonumber\\
& +\big(1+\cos(v_{1})+\cos(v_{3})\big)\big(c(u_{2})-c(u_{6})\big)
\nonumber\\ 
&+\big(1+\cos(v_{2})+\cos(v_{3})\big)\big(c(u_{3})-c(u_{5})\big)
\nonumber\\
& +\big(\sin(v_{1})+\sin(v_{2})\big)\big(s(u_{1})+s(u_{4})\big)
\nonumber\\
& -\big(\sin(v_{1})-\sin(v_{3})\big)\big(s(u_{2})+s(u_{6})\big)
\nonumber\\
&
-\big(\sin(v_{2})+\sin(v_{3})\big)\big(s(u_{3})+s(u_{5})\big)\ ,
\end{align}
where $v_{1}=\frac{2}{3}\pi(\nu_{1}+3\nu_{2}+4)$,
$v_{2}=\frac{2}{3}\pi(2\nu_{1}+3\nu_{2}+5)$, 
$v_{3}=\frac{2}{3}\pi(\nu_{1}+1)$ and 
$s(x)=\sin\big(\frac{2\pi x}{3\kappa}\big)$. (The $u_i$ are defined as
in (\ref{abb}).) This is easily seen to
agree with the right hand side of \eqref{eq:killer1}. 

Let $\rho(\mathcal{B}_{k}^\omega)
=\rho_{0}(\mathcal{B}_{k}^\omega)\,\cup\,
\rho_{1}(\mathcal{B}_k^\omega)\,
\cup \, \rho_{2}(\mathcal{B}_{k}^\omega)$ and
$\mathcal{R}_{k}=\mathcal{P}_{k+2}\setminus
\rho(\mathcal{B}_{k}^\omega)$. The special elements 
$\nu \in \mathcal{P}_{k+2}$ in (\ref{eq:killer1})
which satisfy $\nu_{1}=2$ mod $3$ are
precisely the images $\nu=\tilde{\mu}$ under 
\eqref{eq:idd} of a suitable element $\mu$ of
$\mathcal{E}^{\omega}_{k}$. The key relation $\eqref{eq:killer1}$, 
together with~\eqref{eq:nim}, therefore implies that the $D_{4}$
NIM-rep can be written as a sum of $G_{2}$ fusion matrices, 
\begin{align}
\mathcal{N}_{\lambda a}^{\phantom{\lambda a}b} \ 
& =\ \sum_{\gamma}\varphi^{\phantom{\lambda}\gamma}_{\lambda}
\sum_{\mu \in \mathcal{P}_{k+2}} S^{\prime}_{a\mu}\, 
\frac{S^{\prime}_{\gamma\mu}}{S^{\prime}_{0\mu}}\,
\Big(S^{\prime*}_{\rho_{0}(b)\, \mu}+S^{\prime*}_{\rho_{1} (b)\, \mu}
-S^{\prime*}_{\rho_{2} (b)\, \mu}\Big)\nonumber\\
&=\ \sum_{\gamma}\varphi^{\phantom{\lambda}\gamma}_{\lambda}\,
\Big(N_{\gamma a}^{\phantom{\gamma a} \rho_{0}(b)}
+N_{\gamma a}^{\phantom{\gamma
a}\rho_{1} (b)}-N_{\gamma a}^{\phantom{\gamma
a}\rho_{2} (b)}\Big)\nonumber \\
&=\ \sum_{i=0}^{2}\sum_{\gamma}\varepsilon_{i}\,
\varphi^{\phantom{\lambda}\gamma}_{\lambda}N_{\gamma
a}^{\phantom{\gamma a}\rho_{i}(b)} \ ,
\label{D4NIMrep}
\end{align}
where $N_{\gamma}$ denote $G_{2}$ fusion matrices at level $k+2$ and
$\varepsilon_{i}$ accounts for the signs. [Explicitly
$\varepsilon_0=\varepsilon_1=+1$ and $\varepsilon_2=-1$.]  This is the
relation~\eqref{NIMrep} we proposed in section 1.1.  Note
that~\eqref{D4NIMrep} is valid for all highest weights $\lambda$ of
$D_{4}$, not only for the ones appearing in $\mathcal{P}_{k}^{+}
(D_{4})$. In fact we can continue $\mathcal{N}_{\lambda}$ and
$N_{\gamma}$ outside of the usual domain by rewriting the ratios of
$S$-matrices appearing in~\eqref{eq:nim} as characters of the finite Lie
algebras.

According to the argument given in section 1.1, there is only one
further ingredient we need to show.
This concerns the behaviour of the $G_{2}$-Weyl dimensions under the 
maps $\rho_{i}$, and is summarised in the
relations~\eqref{keydim1} and~\eqref{keydim2}. Thus we need to prove
that 
\begin{alignat}{2} \label{eq:affdim1}
\textrm{dim}_{G_{2}}(\rho_{i}(b))
\ &=\ \varepsilon_{i}\,\textrm{dim}_{G_{2}}(b) 
&\qquad & \textrm{mod}\; M(G_{2},k+2) \ , \\
\intertext{and that any element $r \in \mathcal{R}_{k}$ satisfies}
\label{eq:fehl}
\textrm{dim}_{G_{2}}(r)\ &=\ 0 &  &
\textrm{mod}\; M(G_{2},k+2)\ .
\end{alignat}
The dimension of a $G_{2}$ representation $(b_{1},b_{2})$ is given by 
\[
\textrm{dim}_{G_{2}}(b_{1},b_{2})\ =\ \frac{1}{120}
(b_{1}+1)(b_{2}+1)(b_{1}+b_{2}+2)
(b_{1}+2b_{2}+3)(b_{1}+3b_{2}+4)(2b_{1}+3b_{2}+5)\  .
\]
To prove \eqref{eq:affdim1} we find by explicit computation that 
\begin{equation} \label{eq:dimtrick}
\textrm{dim}_{G_{2}}(\rho_i(b))
\ =\ \varepsilon_i\, \textrm{dim}_{G_2}(b)
+\frac{M(G_2,k+2)}{F}\;p^{5}(b) \  ,
\end{equation} 
where $p^{5}(b)$ denotes a $k$ and $\rho_{i}$-dependent polynomial 
of order $5$ in the variables $b_{1},b_{2}$ with integer coefficients,
and $F=\frac{120}{\textrm{gcd}(k+6,60)}$. Thus it remains to show that  
$\frac{p^{5}}{F}$ is an integer. If $8$ does not divide 
$k+6$, then $M(G_2,k+2)$ and $F$ are coprime. Since
$\frac{M(G_2,k+2)}{F}p^{5}$ is  an integer, $\frac{p^{5}}{F}$
has to be an integer as well and we are done. If $8$ is a divisor of
$k+6$, then $F$ and $M(G_2,k+2)$ have greatest common divisor $2$. The
result then follows provided that $p^{5}$ is even, which is
easily verified.

To show~\eqref{eq:fehl} we first have to identify the elements of
$\mathcal{R}_{k}$. It is convenient to write this set as the
(disjoint) union of the two subsets $\mathcal{R}^1_{k}$ and 
$\mathcal{R}^2_{k}$. The first of them is defined by 
\begin{equation}
\mathcal{R}^1_{k} = \{(b_{1},b_{2}) \in \mathcal{P}_{k+2}\, \vert \,
(b_{1},b_{2})=(k+2-3j,j), \; j \in \mathbb{N}_{0}\}\  .
\end{equation}
The set $\mathcal{R}^2_{k}\equiv \mathcal{R}_{k} \setminus
\mathcal{R}^1_{k}$ depends in a more complicated manner on $k$. To
describe it explicitly we therefore distinguish the three cases:
\begin{itemize}
\item $k=0$ mod $3$\\ 
\hspace*{1cm} 
$\mathcal{R}^2_{k}= \{(2+3j,k/3-1-2j) \in \mathcal{P}_{k+2}, 
\; j \in \mathbb{N}_{0}\}$
\item $k=1$ mod 3\\ 
\hspace*{1cm} 
$\mathcal{R}^2_{k}=\{(0,(k+2)/3)\}\cup\{(1+3j,(k-1)/3-2j) \in
\mathcal{P}_{k+2}, \; j \in \mathbb{N}_{0}\}$ 
\item $k=2$ mod 3 \\
\hspace*{1cm}
$\mathcal{R}^2_{k}=\{(0,(k+1)/3)\}\cup\{(3+3j,(k+1)/3-2(j+1))\in
\mathcal{P}_{k+2} , \; j \in \mathbb{N}_{0}\}\ .$ 
\end{itemize}
For any $r \in \mathcal{R}_{k}$ one then easily checks that 
\begin{equation}
\textrm{dim}_{G_2}(r)\ =\ \frac{M}{F}p^5(j)\  
\end{equation}
with some polynomial $p^{5}$ in $j$ of order 5.  
One finds that the polynomials $p^5(j)$ are even 
whenever 8 divides $k+6$. Using the same arguments as above, this then 
finishes the proof of~\eqref{eq:fehl}. 
It remains to check that we have identified the complete set
$\mathcal{R}_{k}$ correctly. Because
$\rho_{i}(\mathcal{B}_{k}^\omega)\,\cap\,
\rho_{j}(\mathcal{B}_{k}^\omega)=\emptyset$ for $i\neq j$ 
we have $|\rho(\mathcal{B}_{k}^\omega)| =3 \,|\mathcal{B}_{k}^\omega|$.  
In order to see that
$\mathcal{P}_{k+2}=\rho(\mathcal{B}_{k}^\omega)\cup \mathcal{R}_{k}$,
it is therefore sufficient to count the number of elements of the
different sets. One easily finds 

\renewcommand{\arraystretch}{1.8}
$$
|\mathcal{P}_{k+2} | = \left\{
\begin{array}{ll} 
\frac{1}{4}(k+4)^2 & \quad k \; \textrm{even}\\
\frac{1}{4}(k+3)(k+5) & \quad k \; \textrm{odd}\\
\end{array}\right.
$$
as well as 
$$
|\mathcal{R}_{k} | = \left\{
\begin{array}{llr} 
\frac{1}{2}(k+2) & \quad k=0 & \,\text{mod}\  6 \\
\frac{1}{2}(k+4) & \quad k=2,4 & \text{mod}\  6\\
\frac{1}{2}(k+3) & \quad k=3 & \text{mod} \ 6 \\
\frac{1}{2}(k+5) & \quad k=1,5 & \text{mod}\  6\\
\end{array}\right.
\qquad
|\mathcal{B}_{k}^{\omega } | = \left\{
\begin{array}{llr} 
\frac{1}{12}k^{2}+\frac{1}{2}k+1 & \quad k=0 &\,\text{mod}\  6 \\
\frac{1}{12}(k+2)(k+4) & \quad k=2,4 & \text{mod}\  6\\
\frac{1}{12}(k+3)^2 & \quad k=3 & \text{mod} \ 6 \\
\frac{1}{12}(k+1)(k+5)& \quad k=1,5 & \text{mod}\  6 \\  
\end{array}\right.
$$
\renewcommand{\arraystretch}{1} 

\noindent Using these formulae it is then easy to show that 
$|\mathcal{P}_{k+2}|=|\rho(\mathcal{B}_{k}^\omega)|\,+\,
|\mathcal{R}_{k}|$. 
This completes the proof.

\subsection{Uniqueness}
It remains to prove that the solution we found is unique up to an
overall rescaling of the charge. To this end we show that any
solution of the charge constraint modulo some integer $M'$ satisfies
the relation
\begin{equation}\label{uniqueness}
q_a\ =\ \dim  (a)\, q_0 \quad \mod M' \  ,
\end{equation}
and thus is obtained from our solution (\ref{eq:choice}) by scaling
with the factor $q_{0}$.

To prove~\eqref{uniqueness} we first want to show that any $G_{2}$
representation $a$ can be obtained as restriction of a linear
combination of $D_4$ representations $\lambda_j$ with integer
coefficients $z_j$.  We explicitly allow negative multiplicities and
write formally
\[
a \ =\ \bigoplus_j z_j \lambda_j\big|_{G_2}  \ .
\]
Obviously it is sufficient to prove this for the fundamental
representations. Looking at the branching rules~\eqref{D4branching} we
see that the representation $(1,0)$ appears in the decomposition of
$(1,0,0,0)$, so we can write
\[
(1,0)\ =\ \big( (1,0,0,0) - (0,0,0,0) \big) \big|_{G_2} \ .
\]
Similarly, we can express $(0,1)$ as a restriction because it appears
exactly once in the decomposition of $(0,1,0,0)$ together only with
$(1,0)$ (see~\eqref{D4branching}).

Now consider a boundary state labelled by $a$.
We can use~\eqref{D4NIMrep} to write\footnote{Note that the charge
constraint~\eqref{eq:main} as well as the expression~\eqref{D4NIMrep}
for the NIM-rep is valid also for highest weights $\lambda$ which are
not in $\mathcal{P}_{k}^{+} (D_{4})$.} 
\begin{align}
\nonumber
\dim_{G_2}(a)\, q_0\ &=\ \sum_j z_j\, \dim_{D_4} (\lambda_j)\,
q_0\\
\nonumber
&=\ \sum_{j,b} z_j \, \mathcal{N}_{\lambda_j 0}{}^b q_b \qquad \mod M'\\
\nonumber
&=\ \sum_{i,j,\gamma,b} z_j \, \varepsilon_i \,
\varphi_{\lambda_j}{}^{\gamma} \, N_{\gamma 0}{}^{\rho_i(b)}\,  q_b\\
\nonumber
&=\ \sum_{i,b} \varepsilon_i\, N_{a0}{}^{\rho_i(b)} \, q_b\\
&=\ q_a \ .
\end{align}
In the last step we used the fact that $\rho_{i}
(\mathcal{B}_{k}^\omega)$ and $\mathcal{B}_{k}^\omega$ are disjoint
for $i\not= 0$, so that only $i=0$ contributes. This concludes the
proof of~\eqref{uniqueness}.  

\section{The $\boldsymbol{E_{6}}$ case with charge conjugation} 
The analysis for the case of $E_6$ is fairly similar to the $D_4$ case
discussed in the previous section, and we shall therefore be somewhat
briefer. For $E_{6}$ the invariant subalgebra under charge
conjugation is $\bar{\mathfrak{g}}^{\omega}=F_{4}$. Again we have the
identity  
\begin{equation}
M(E_{6},k)\ =\ M(F_{4},k+3)\  =\  \frac{ k+ 12}{{\rm gcd}(k+12,2^3 \cdot
3^2\cdot 5 \cdot 7 \cdot 11)} \ .
\end{equation}
As before we therefore expect that 
\begin{equation} \label{eq:main2}
\textrm{dim}_{E_{6}}(\lambda)\,q_a \ =\ 
\sum_{b\in \mathcal{B}_{k}^\omega}
\mathcal{N}_{\lambda a}^{\phantom{\lambda a}b} \, 
q_b \qquad \textrm{mod} \;M(F_{4},k+3) \ ,
\end{equation}
where 
\begin{equation} \label{eq:choice2}
q_b \ =\ \textrm{dim}_{F_4}(b) \ . 
\end{equation}
The order 2 automorphism $\omega$ of $E_{6}$ maps the Dynkin labels
$(\mu_{0};\mu_{1},\mu_{2},\mu_{3},\mu_{4},\mu_{5},\mu_{6})$ to
$(\mu_{0};\mu_{5},\mu_{4},\mu_{3},\mu_{2},\mu_{1},\mu_{6})$, and thus 
the set of exponents is 
\begin{equation}
\mathcal{E}^{\omega}_{k}
\ =\ \{(\mu_{0};\mu_{1},\mu_{2},\mu_{3},\mu_{2},\mu_{1},\mu_{6})
\in \mathbb{N}_{0}^{7} \, \vert \,
\mu_{0}+2\mu_{1}+4\mu_{2}+3\mu_{3}+2\mu_{6}=k\}\ .
\end{equation}
The twisted algebra
here is $E_{6}^{(2)}$.  The set of boundary states at level $k$ is 
explicitly given by
$\mathcal{B}_{k}^\omega=\{(b_{0};b_{1},b_{2},b_{3},b_{4})\in 
\mathbb{N}_{0}^{5}\,\vert\, b_{0}+2b_{1}+3b_{2}+4b_{3}+2b_{4}=k\}.$ As
in the last section, it is possible to identify the $\psi$-matrix of
$E_6$ at level $k$ with the $S$-matrix of $F_4$ at level $k+3$
(see also \cite{Gaberdiel:2002qa}),
\begin{equation}
\psi_{b\mu}\ =\ 2\,S^{\prime}_{b\tilde{\mu}}\ ,
\end{equation}
where $\tilde{\mu}$ is now defined by 
\begin{equation} \label{eq:idd2}
\mu \ \mapsto\  \tilde{\mu}= 
(\mu_{0};2\mu_{1}+1,2\mu_{2}+1,\mu_{3},\mu_{6})\ .
\end{equation}  
As before we observe that if $\mu\in{\cal E}^\omega_k$, then 
$\tilde\mu\in \mathcal{P}^+_{k+3}(F_4)\equiv \mathcal{P}_{k+3}$,
where the latter is explicitly defined as 
$\mathcal{P}_{k+3}=
\{(\tilde\mu_{1},\tilde\mu_{2},\tilde\mu_{3},\tilde\mu_{4})\in
\mathbb{N}^{4}_{0}\,\vert\, \tilde\mu_{1}+2\tilde\mu_{2}
+3\tilde\mu_{3}+2\tilde\mu_{4}\leq k+3\}$.
Furthermore, we can express ratios of $S$-matrices of $E_{6}$ by those
of $F_{4}$, 
\[
\frac{S_{\lambda\mu}}{S_{0\mu}}
\ =\ \sum_{\gamma}\varphi^{\phantom{\lambda}\gamma}_{\lambda}\ 
\frac{S^{\prime}_{\gamma\tilde{\mu}}}{S^{\prime}_{0\tilde{\mu}}} 
\ .
\]
Here $S$ denotes the $E_{6}$ $S$-matrix at level $k$, $S^{\prime}$
is the $F_{4}$ $S$-matrix at level $k+3$, and
$\varphi^{\phantom{\lambda}\gamma}_{\lambda}$ describes the branching 
$E_6\supset F_4$; for the six fundamental representations of $E_6$
the branching rules are 
\begin{eqnarray}
(1,0,0,0,0,0) & \simeq & (0,0,0,0,1,0)  \, \to\,   
(1,0,0,0) \oplus (0,0,0,0) \nonumber \\
(0,1,0,0,0,0) & \simeq & (0,0,0,1,0,0) \, \to\,  
(0,1,0,0) \oplus (0,0,0,1)\oplus (1,0,0,0) \nonumber \\
(0,0,1,0,0,0) & \to & (0,0,1,0) \oplus (1,0,0,1) 
\oplus 2\cdot (0,1,0,0) \oplus (0,0,0,1)
\nonumber \\
(0,0,0,0,0,1) & \to & (0,0,0,1) \oplus (1,0,0,0) \ .
\label{E6branching}
\end{eqnarray}
The relevant affine mappings are in this case
\begin{equation} \label{eq:aff2}
\begin{array}{l}
\rho_0(b)\ =\ (b_{1},b_{2},b_{3},b_{4})\\
\rho_1(b)\ =\ (k-2b_{1}-3b_{2}-4b_{3}-2b_{4},1+b_{1}+b_{2},b_{3},b_{4})\\
\rho_2(b)\ =\ (k+1-b_{1}-3b_{2}-4b_{3}-2b_{4},b_{2},b_{3},b_{4})\\
\rho_3(b)\ =\ (k-2b_{1}-3b_{2}-4b_{3}-2b_{4},b_{1},b_{2}+b_{3}+1,b_{4}) \ , 
\end{array}
\end{equation}
which map boundary states at level $k$ to dominant weights of $F_{4}$
at level $k+3$, {\it i.e.} to elements of $\mathcal{P}_{k+3}$. 
There is a similar identity to \eqref{eq:killer1} for the $S$-matrices 
\begin{equation} \label{eq:killer2}
S^{\prime}_{\rho_{0}(b)\, \nu}
+S^{\prime}_{\rho_{1}(b)\, \nu}
-S^{\prime}_{\rho_{2}(b)\, \nu}
-S^{\prime}_{\rho_{3}(b)\, \nu} \ = \ \left\{ \begin{array}{ll} 4 
\, S^{\prime}_{b\, \nu} \qquad & \textrm{if $\nu_{1}=\nu_{2}=1$ mod $2$}
\\ 0 
\qquad & \textrm{otherwise,} \end{array} \right. 
\end{equation}
where $b\in \mathcal{B}_{k}^\omega$ and $\nu \in
\mathcal{P}_{k+3}$. Again, the elements which satisfy
$\nu_{1}=\nu_{2}=1$ mod $2$ are 
precisely the images $\nu=\tilde{\mu}$ of an element $\mu$ of
$\mathcal{E}_k^{\omega}$ under the mapping \eqref{eq:idd2}. 
By $\rho(\mathcal{B}_{k}^\omega)$ we 
denote the union of the images of $\mathcal{B}_{k}^\omega$ under the
maps $\rho_{i}$, $\rho (\mathcal{B}_{k}^\omega)=
\bigcup_{i=0}^{3}\rho_{i} (\mathcal{B}_{k}^\omega)$. 
The elements of $\mathcal{P}_{k+3}$ which are not reached by the maps
form the set $\mathcal{R}_{k}=\mathcal{P}_{k+3}\setminus 
\rho(\mathcal{B}_{k}^\omega)$.   

Using essentially the same arguments as for the case of $D_4$
discussed in the last section, we can then show that the $E_{6}$
NIM-rep can be expressed in terms of $F_{4}$ fusion matrices as 
\begin{equation}\label{NIMasfus2}
\mathcal{N}_{\lambda a}^{\phantom{\lambda a} b} 
\ =\ \sum_{\gamma}\varphi^{\phantom{\lambda}\gamma}_{\lambda}\,
\Big(N_{\gamma a}^{\phantom{\gamma a}\rho_{0}(b)}
+N_{\gamma a}^{\phantom{\gamma 
a}\rho_1(b)}-N_{\gamma a}^{\phantom{\gamma
a}\rho_2(b)}-N_{\gamma a}^{\phantom{\gamma
a}\rho_3(b)}\Big)\ =\ \sum_{i=0}^{3}\sum_{\gamma}\varepsilon_{i}\,
\varphi^{\phantom{\lambda}\gamma}_{\lambda}N_{\gamma
a}^{\phantom{\gamma a}\rho_{i}(b)} \ , 
\end{equation}
where the $\varepsilon_{i}$ account for the signs. [Explicitly, 
$\varepsilon_1=\varepsilon_2=+1$ and
$\varepsilon_3=\varepsilon_4=-1$.] 
Following the argument of section 1.1, it thus only remains to show 
that 
\begin{alignat}{2} \label{eq:affdim2}
\textrm{dim}_{F_{4}}(\rho_{i}(b))
\ &=\ \varepsilon_{i}\,\textrm{dim}_{F_{4}}(b)& \qquad& 
\textrm{mod}\; M(F_{4},k+3)  \ , \\ 
\intertext{and for all $r \in \mathcal{R}_{k}$}
\label{eq:fehl2}
\textrm{dim}_{F_4}(r)\ &=\ 0  & \qquad & 
\textrm{mod}\; M(F_4,k+3) \ .
\end{alignat}
To prove equation \eqref{eq:affdim2} we note that 
\begin{equation}
\textrm{dim}_{F_{4}}(\rho_{i}(b))\ =\ 
\varepsilon_{i}\,\textrm{dim}_{F_{4}}(b)
+\frac{M(F_4,k+3)}{F}\, p^{23}(b)
\end{equation}
where 
\begin{equation}
F\ =\ \frac{2^{15}\cdot 3^7\cdot 5^4\cdot 7^2\cdot
11}{\textrm{gcd}(k+12,2^3\cdot 3^2\cdot 5\cdot 7\cdot 11)}  
\end{equation}
and $p^{23}$ is a $k$ and $\rho_{i}$-dependent polynomial (with
integer coefficients) of degree $23$ in the labels $b_i$. Now
$M(F_4,k+3)$ and $F$ are coprime whenever $2^4$, $3^3$, $5^2$ and
$7^2$ do not divide $k+12$; in this case \eqref{eq:affdim2} is proven
as before. Otherwise the analysis is more involved and many cases
would have to be distinguished. We have not attempted to analyse all
of them in detail, but we have performed a numerical check up to
fairly high levels. This seems satisfactory, given that the identities
for $M(E_6,k)$ and $M(F_4,k)$ have also only be determined
numerically.

Finally, we need to show the identity~\eqref{eq:fehl2}. This requires
a good description of the set $\mathcal{R}_{k}$. Here it is convenient
to write it as the union of four disjoint subsets which are defined by
\renewcommand{\arraystretch}{1.6}
\begin{align*}
\mathcal{R}_{k}^1\ &=\ \left\{ \begin{array}{p{11.2cm}l} 
$\{b \in \mathcal{P}_{k+3}\,\vert\, 
b=(1+2j_1,j_2,j_3,(k+2)/2-j_1-j_2-2j_3) 
\} $ & k \; \textrm{even} \\ $\{b \in \mathcal{P}_{k+3}\,\vert\, 
b=(2j_1,j_2,j_3,(k+3)/2-j_1-j_2-2j_3)\}$ & k \; \textrm{odd}\\ 
\end{array} \right.\\
\mathcal{R}_{k}^2\ &=\ \left\{ \begin{array}{p{11.2cm}l} $ \{b 
\in \mathcal{P}_{k+3}\,\vert\, b=(2j_1,2j_2,j_3,
(k+2)/2-j_1-3j_2-2j_3) 
\}$ & k \; \textrm{even} \\ 
$\{b \in \mathcal{P}_{k+3}\,\vert\, 
b=(1+2j_1,2j_2,j_3,(k+1)/2-j_1-3j_2-2j_3)\}$ & k \; \textrm{odd}\\ 
\end{array} \right.\\
\mathcal{R}_{k}^3\ &=\ \left\{ \begin{array}{p{11.2cm}l} $\{b \in 
\mathcal{P}_{k+3}\,\vert\, b=(1+2j_1,1+2j_2,j_3,
(k-2)/2-j_1-3j_2-2j_3) 
\} $ & k \; \textrm{even} \\ $ \{b \in \mathcal{P}_{k+3}\,\vert\, 
b=(2j_1,1+2j_2,j_3,(k-1)/2-j_1-3j_2-2j_3)\} $& k \; \textrm{odd}\\ 
\end{array} \right.\\
\mathcal{R}_{k}^4\ &=\ \left\{ \begin{array}{p{11.2cm}l} $\{b \in 
\mathcal{P}_{k+3}\,\vert\, b=(j_1,1+2j_2,j_3,
(k-2)/2-j_1-3j_2-2j_3)
 \} $ & k \; \textrm{even} \\ $\{b \in \mathcal{P}_{k+3}\,\vert\, 
b=(j_1,2j_2,j_3,(k+1)/2-j_1-3j_2-2j_3)\} $& k \; \textrm{odd,}\\ 
\end{array} \right.
\end{align*}
where $(j_1,j_2,j_3) \in \mathbb{N}_{0}^3$. The same arguments as
before show that the elements $r$ in these sets satisfy
$\textrm{dim}_{F_4}(r)=0$ $\textrm{mod}\; M(F_4,k+3)$. Again, this is
proven only if $k+12$ is not divisible by $2^4,3^3,5^2$ or $7^2$; for
the other levels we have only performed numerical checks. 

Finally, by counting the elements of the different sets we can confirm
(as before) that we have correctly identified the set 
$\mathcal{R}_k$. This completes the proof for the case of $E_6$. 

\subsection{Uniqueness} The proof of uniqueness is analogous to the
$D_{4}$ case. It only remains to show that all fundamental
representations of $F_{4}$ can be obtained as restrictions of $E_{6}$
representations. From the branching rules~\eqref{E6branching} we see
immediately that this is true for $(1,0,0,0)$, $(0,0,0,1)$ and
$(0,1,0,0)$. The remaining fundamental representation $(0,0,1,0)$
appears in the decomposition of $(0,0,1,0,0,0)$, but it comes together
with $(1,0,0,1)$. The latter representation can be obtained from the
other fundamentals by the $F_{4}$ tensor product 
\[
(1,0,0,0)\otimes (0,0,0,1)\ \to \ 
(1,0,0,1)\oplus (1,0,0,0)\oplus (0,1,0,0)\ .
\]
Hence, also $(0,0,1,0)$ can be written in terms of the restriction of 
$D_{4}$-representations. 
\bigskip

\noindent {\it Note added:} While we were in the process of writing up
  this paper we became aware of \cite{MV} which contains closely
  related work.

\bigskip 

\noindent 
{\bf Acknowledgements:} This research has been partially supported by 
the Swiss National Science Foundation and the Marie Curie network
`Constituents, Fundamental Forces and Symmetries of the Universe'
(MRTN-CT-2004-005104). The work of S.F.\ was supported by the Max
Planck Society and the Max Planck Institute for Gravitational Physics
in Golm. We thank Terry Gannon for useful communications. This paper
is largely based on the Diploma thesis of T.M.\ \cite{TM}.

\providecommand{\href}[2]{#2}\begingroup\raggedright
\endgroup

\end{document}